\documentclass[%
 reprint,
 amsmath,amssymb,
 aps,
floatfix
]{revtex4-2}

\usepackage{amssymb}
\usepackage{amsmath}
\usepackage{bm}
\usepackage[dvipsnames]{xcolor}
\usepackage[normalem]{ulem}

\usepackage{hyperref}
\usepackage{graphicx}
\usepackage{mathtools}

\usepackage{etoolbox}
\usepackage{tikz}
\newrobustcmd*{\mysquare}[1]{\tikz{\filldraw[draw=#1,fill=#1] (0,0)
rectangle (0.2cm,0.2cm);}}

\begin{document}


\title{Growth regimes in three-dimensional phase separation of liquid-vapor systems}

\author{G. Negro}
\affiliation{Dipartimento di Fisica, Universit\'a degli Studi di Bari and INFN, Sezione di Bari, via Amendola 173, Bari, I-70126, Italy}
\author{G. Gonnella}
\affiliation{Dipartimento di Fisica, Universit\'a degli Studi di Bari and INFN, Sezione di Bari, via Amendola 173, Bari, I-70126, Italy}
\author{A. Lamura}
\email[Corresponding author, E-mail: ]{antonio.lamura@cnr.it}
\affiliation{Istituto Applicazioni Calcolo, CNR, Via Amendola 122/D, I-70126 Bari, Italy}
\author{S. Busuioc}
\affiliation{Institute for Advanced Environmental Research, West University of Timi\c soara, Bd. Vasile P\^ arvan 4, 300223 Timi\c soara, Romania}
\author{V. Sofonea}
\affiliation{Center for Fundamental and Advanced Technical Research, Romanian Academy, Bd. Mihai Viteazul 24, 300223 Timi\c soara, Romania}
\begin{abstract}
The liquid-vapor phase separation is investigated via lattice Boltzmann simulations in three dimensions. 
After expressing length and time scales in reduced physical units, we combined data from several large simulations (on $512^3$ nodes), with different values of viscosity, surface tension and temperature, to obtain a single curve of rescaled length $\hat{l}$ as a function of rescaled time $\hat{t}$.
We find evidence of the existence of \textcolor{black}{kinetic} and inertial regimes with growth exponents $\alpha_d=1/2$ and $\alpha_i=2/3$ over several time decades, with a crossover from 
$\alpha_d$ to $\alpha_i$ at $\hat{t} \simeq 1$. 
\textcolor{black}{This allows us to rule out the existence of a viscous regime with $\alpha_v=1$ in three-dimensional liquid-vapor isothermal phase separation, differently from what happens in binary fluid mixtures.}
An in-depth analysis of the kinetics of the phase separation process, as well as a characterization of the morphology and the flow properties, are further presented in order to provide clues into the dynamics of the phase-separation process.

\end{abstract}
\maketitle

\section{Introduction}

In fluid mixtures, the interplay of hydrodynamics and thermodynamics gives rise to a wealth of complex effects of great fundamental significance and industrial practical value \cite{onuki2002}. One of the most striking examples is provided by phase separation. When a fluid is brought into a coexistence region by a quench from a temperature above the critical point, the initial single phase becomes unstable and the phase separation occurs through the formation at a molecular length scale of domains that grow in time to a macroscopic extent. 

This phenomenon has been extensively studied in the past years by theoretical approaches, experiments, and numerical simulations \cite{bray1994,onuki2002,cates2017}. The observation that domains grow in self-similar ways led to the scaling concept that a typical size $L$ exists in forming domains. This quantity follows a power law $L \sim t^{\alpha}$ where $t$ is the time and the growth exponent $\alpha$ depends only on the physical mechanism operating during phase separation. In the following, we will refer to systems with symmetric composition producing bicontinuous growing patterns. 

In the case of binary fluids, the initial growth is purely diffusive since hydrodynamics does not play a significant role. The growth proceeds via the Lifshitz-Slyozov (LS) mechanism by which smaller droplets shrink by diffusion of material to larger droplets that grow. In this case, the value $\alpha_d=1/3$ is observed \cite{bray1994}. Hydrodynamics is, in general, relevant at late times and the coupling with the velocity field is responsible for changing the value of the growth exponent. In particular, two regimes can be accessed which are characterized by viscous and inertial hydrodynamic growths with exponents  $\alpha_v=1$ and $\alpha_i=2/3$, respectively\cite{bray1994}. \textcolor{black}{The expected mechanism leading to the exponent $\alpha_v=1$ }
\textcolor{black}{in three dimensions} is the Rayleigh instability, which produces the pinch off of the necks of connected domains. The resulting broken channels retract promoting the growth of domains. \textcolor{black}{It is worth mentioning that the exponent $\alpha_v=1$ has also been found in two dimensions for binary fluids with initial asymmetric droplet
morphology} \cite{wagner01}. The inertial regime, \textcolor{black}{with $\alpha_i=2/3$}, is driven by the Laplace pressure difference between the inner and outer parts of a domain producing more spherical patterns. \textcolor{black}{The viscous and the inertial regimes were observed in  simulations of three-dimensional binary fluid mixtures \cite{kendon1999,kendon2001}, providing a convincing confirmation of the existence of these scaling regimes.}

\begin{table}
\textcolor{black}{
\caption{\textbf{Growth exponents}. The table summarizes the 3D growth exponents and their different mechanisms \textcolor{black}{for isothermal one-component nonideal fluids}.}\label{table:exponents}
\centering
\begin{tabular}{ |l | c|}
\hline\hline
 \bf{Growth Exponent}  & \bf{Mechanism}\\
 \hline 
$1/3$ & Lifshitz-Slyozov \\
\hline  
$1/2$ & \textcolor{black}{Kinetic evaporation-condensation} \\
\hline 
 $1$ &   \begin{tabular}[c]{@{}l@{}}Viscous hydrodynamic regime \\ (Rayleigh instability )\end{tabular}\\
 \hline 
 $2/3$ &   \begin{tabular}[c]{@{}l@{}}Inertial hydrodynamic regime \\ (Laplace pressure difference) \end{tabular}\\
\hline\hline
\end{tabular}}
\end{table}

\textcolor{black}{The study of phase separation in liquid-vapor systems did not receive the same attention. 
When hydrodynamics is absent, the growth dynamics is found to have an exponent  $\alpha_d = 1/2$ for isothermal one-component nonideal fluids~\cite{Koch1983}  and has been observed in molecular dynamics (MD) simulations both in two and three dimensions \cite{yamamoto1994}. 
In this case the growth exponent can be attributed to a different realization of the LS evaporation-condensation mechanism, where the transport of molecules is kinematic rather than diffusive \cite{onuki2002,Koch1983,tateno2021}.
It was recently put forward \cite{tateno2021} that this coarsening law is of more general validity for liquid-vapor-type phase separation in simple and complex fluids, far from the critical point. 
The same exponent was also found for dynamically symmetric binary mixtures where it was attributed to the Brownian-coagulation mechanism for two-dimensional fluids \cite{san1985}. For two or more component liquid-vapor
systems the overall physics is more complicated since both diffusive and hydrodynamic processes are relevant
due to the different compositions of liquid and vapor phases \cite{ridl18}. }
For completeness we add that the use of stochastic thermostats (e.g., Andersen thermostat \cite{frenkel2002}) in MD studies would give a diffusive growth exponent $1/3$ \cite{das2011}, as in the LS mechanism.
\textcolor{black}{Table \ref{table:exponents} offers a concise overview of all the growth exponents discussed thus far, along with their corresponding physical mechanisms.}

\textcolor{black}{An interesting question, concerning phase separation of liquid-vapor systems, is whether the same hydrodynamic growth exponents of binary mixtures are also valid.} Previous numerical studies in two dimensions, based on the lattice Boltzmann (LB) algorithm, found the value $\alpha_d= 1 /2$ at high viscosity and $\alpha_i = 2/3$ at low viscosity \cite{osborn1995,swift1996,sofonea2004,xu2011,zhang2019}. A different numerical approach was adopted in Ref.~\cite{lamorgese2009}, where the Navier-Stokes equation, supplemented by a nonequilibrium body force to describe a van der Waals fluid, was numerically solved in three dimensions. In Ref.~\cite{lamorgese2009}, only the value $\alpha_i = 2/3$ was observed and it was argued that this behavior is due to the balancing of capillary forces, active at initial times, with inertial forces acting at late times. For the sake of completeness, we add that in three-dimensional molecular dynamics simulations \cite{das2011} it was found, after the initial growth, a narrow time window with an \textcolor{black}{asymptotic}
growth exponent \textcolor{black}{oscillating around the mean value $1$. However, this result was obtained
on a rather small system ($96^3$) for a single case}. 

To the best of our knowledge, the viscous exponent was never observed and a systematic investigation of the phase separation in liquid-vapor systems is still missing. In this work, we present a full numerical study of the quench of a van der Waals fluid below the critical temperature $T_c$ in three dimensions to provide a detailed study of the growth exponents. To this purpose, we adopt a lattice Boltzmann equation (LBE) for liquid-vapor systems based on a Gauss-Hermite projection of the corresponding continuum Boltzmann equation \cite{shan2006}. A body force, derived from a proper free-energy, is added to the LBE \cite{coclite2014,kaehler2015} and the equilibrium distribution functions are properly redefined \cite{Guo2002} to model a van der Waals fluid described by the Navier-Stokes equations. By adopting the approach proposed for binary fluids \cite{kendon2001}, we construct characteristic space $L_0=\eta^2/\rho\sigma$ and time $t_0=\eta^3/\rho\sigma^2$ scales where $\eta$ is the viscosity, $\sigma$ is the surface tension, and $\rho$ is the density of the fluid. By defining dimensionless quantities $\hat{l}=L/L_0$ and $\hat{t} \propto t/t_0$, data from several simulations with different values of $\eta$, $\sigma$, and $\rho$ can be combined to obtain a single $(\hat{l},\hat{t})$ curve. We find evidence for  the growth exponents
 $\alpha_d=1/2$ and $\alpha_i=2/3$ over several time decades with a crossover from one exponent to the other at $\hat{t} \simeq 1$, thus ruling out the existence of a viscous regime with $\alpha_v=1$ in the liquid-vapor phase separation. 

The paper is organized as follows: we will first present the LB model used to simulate a phase-separating liquid-vapor system in three dimensions. We will then discuss the kinetics of the phase separation process quantitatively, and characterize the morphology of the different regimes using the Minkowski functionals \cite{serra1983,mecke1997,mecke1999,xu2011,MICHIELSEN2001461}. Finally, we will discuss the flow properties of the low and high-viscosity regimes. 


\section{Numerical Model}
\begin{figure}[t!]
 \centering
\includegraphics[width=0.96\columnwidth]{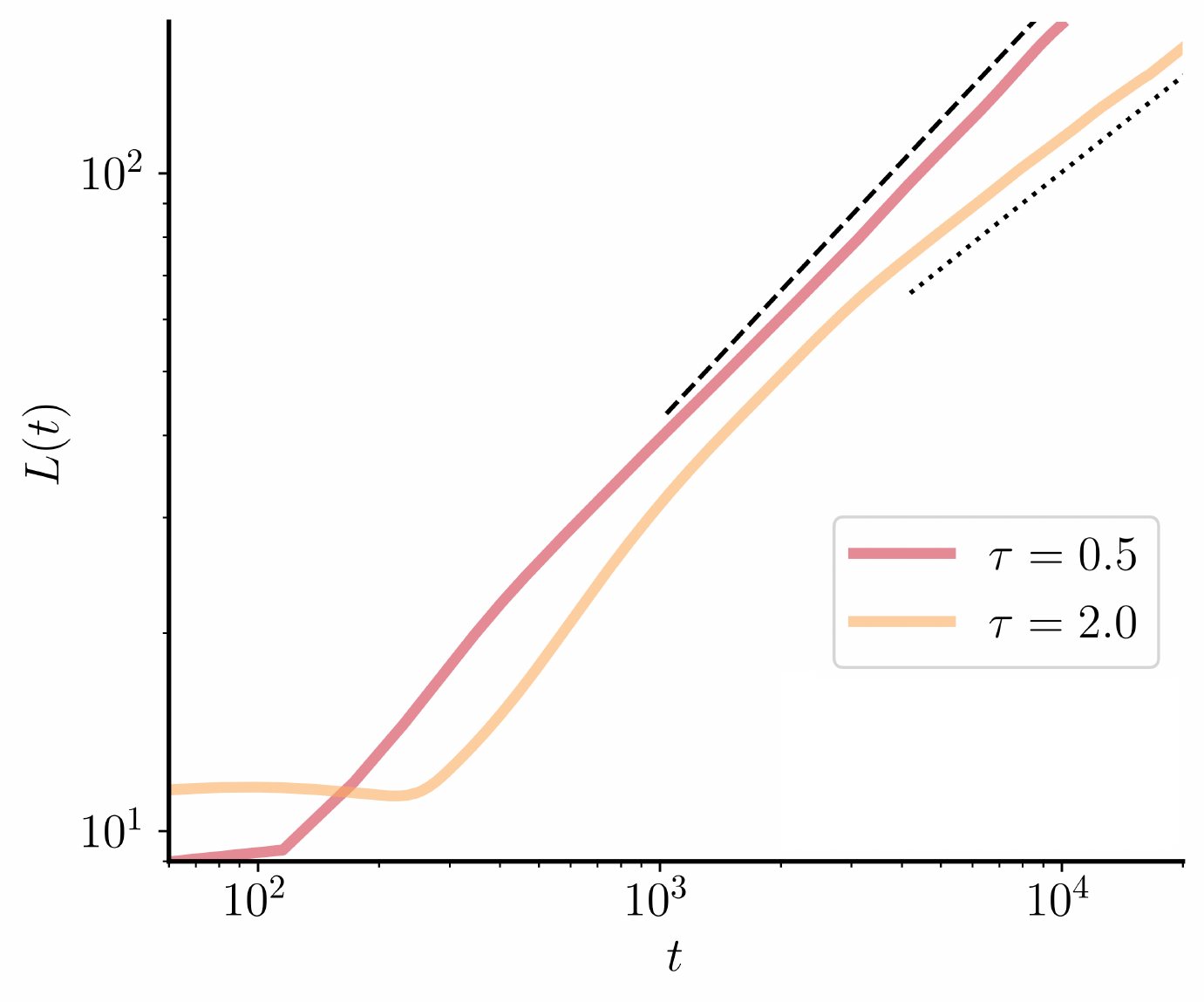}
 \caption{\textbf{Kinetics.} Evolution of domain size $L(t)$, calculated according to Eq.~\eqref{eq:Lt}, for $T=0.95$, $\kappa=0.1$ and two values of the relaxation time, namely $\tau=0.5$ (red) and $\tau=2.0$ (orange). The dashed and dotted black lines have slopes $2/3$ and $1/2$, respectively.
 All quantities are expressed in lattice units.}
 \label{fig:1_radius}
\end{figure}

For the simulation of the liquid-vapor system, in this paper we use the three-dimensional D3Q15 LB model \cite{shan2006,carenzareview} \textcolor{black}{where each site of the lattice is connected to its $6$ first neighbors and its $8$ third neighbors.}
All quantities \textcolor{black}{are expressed}
in lattice units, as discussed in Ref.~\cite{Negro2019}. This isothermal model is defined on a cubic lattice with spacing $\Delta s = 1$ and involves $N=15$ distribution functions ${f_i}$ $(i=0,...,N-1)$ which obey the dimensionless Boltzmann equation
\begin{equation}\label{eq:be}
\begin{split}
 &f_i(\mathbf{r}+\mathbf{e}_i \Delta t,t+\Delta t)-f_i(\mathbf{r}, t)\\= &-\frac{\Delta t}{\tau}\left[f_i(\mathbf{r},t)-f_i^{eq}(\mathbf{r},t)\right]+\Delta{t}\mathcal{F}_i \ , 
\end{split}
\end{equation}
where $\mathbf{r}$ is the position vector in the coordinate space, $t$ is the time, $\{{\mathbf{e}_i}\}\, (i=0,...,N-1)$ is the set of discrete velocities \textcolor{black}{(to be later determined)}, $\Delta t$ is the time step, $\tau$ is the relaxation time, and $\mathcal{F}_i$ is a forcing term to be determined. The moments of the distribution functions $f_i$ define the local fluid density $\rho=\sum_i f_i$ and the local velocity $\mathbf{v}=\sum_i f_i\mathbf{e}_i/\rho$.

The local equilibrium distribution functions ${f_i^{eq}}$ $(i=0,...,N-1)$ are expressed by Maxwell-Boltzmann (MB) distribution. Here we adopt a discretization in velocity space of the MB distribution based on the quadrature of a Hermite polynomial expansion of it. In this way, it is possible to get a LBE that allows us to exactly recover a finite number of leading order of moments on the MB distribution. The form of $f_i^{eq}$ is
\begin{equation}\label{eq:feq}
\begin{split}
 f_i^{eq}(\mathbf{r},t) = &\omega_i\rho\left[1+\mathbf{e}_i\cdot\mathbf{v}+\frac{1}{2}\mathbf{v}\mathbf{v}:(\mathbf{e}_i\mathbf{e}_i-\mathbf{I})\right.\\ 
 &\left.+\frac{1}{2}(T-1)(\mathbf{e}_i\cdot\mathbf{e}_i-3)\right]\, ,
\end{split} 
\end{equation}
where $\mathbf{I}$ is 
the unit matrix and $T$ is the (constant) value of the fluid temperature.
\textcolor{black}{The term in the second line of the equation above, which is proportional to $(T-1)$,
would be absent when simulating isothermal systems at the reference temperature
$T=1$.  
Throughout the present work, we use the critical value $T_c = 1$ as reference and we will consider isothermal liquid-vapor systems at temperatures $T < T_c$.}
This expression of $f_i^{eq}$ allows one to retrieve all the moments of the MB distribution function up to the second order:
\begin{subequations}
\begin{eqnarray}\label{eq:feqmoments}
 \sum_if_i^{eq}&=&\rho \ , \\
 \sum_if_i^{eq}\mathbf{e}_i&=&\rho\mathbf{v}\ , \\
 \sum_if_i^{eq}\mathbf{e}_i\mathbf{e}_i&=&\rho T \mathbf{I}+\rho\mathbf{v}\mathbf{v}\, . 
 \end{eqnarray}
 \end{subequations}
\textcolor{black}{The Cartesian projections of the lattice vectors $\mathbf{e}_i$, which span the velocity space,
are given by the abscissas of the Gauss-Hermite quadrature of order 3 \cite{shan2006}}.
On the chosen lattice $D3Q15$, the previous second order expansion fixes the values $|\mathbf{e}_0|=0$
\textcolor{black}{(rest velocity)}, $|\mathbf{e}_i|=\Delta s/\Delta t=\sqrt{3}$ for $i=1-6$ \textcolor{black}{(first neighbors)}, 
$|\mathbf{e}_i|=\sqrt{3} \Delta s / \Delta t=3$ for $i=7-14$ \textcolor{black}{(third neighbors)}, for the lattice velocities. This requires that the time-step is $\Delta t =\sqrt{3}/3$ with the choice $\Delta s=1$. Finally, the weights at the r.h.s. of Eq.~(\ref{eq:feq}) are $\omega_0=2/9$, $\omega_i=1/9$ for $i=1-6$, $\omega_i=1/72$ for $i=7-14$ \cite{shan2006}. \textcolor{black}{For some cases, we ran  simulations
using the D3Q27 lattice with 27 lattice speeds and found no significant differences in the results presented in the following.}

To simulate a liquid-vapor system we use the forcing scheme proposed in Ref.~\cite{Guo2002}. According to this scheme, the fluid velocity $\mathbf{v}$ is replaced by the physical velocity 
\begin{equation}
 \mathbf{v}^*=\frac{1}{\rho}\sum_if_i\mathbf{e}_i+\frac{1}{2\rho}\mathbf{F}\Delta t \ , 
\end{equation}
where $\mathbf{F}$ is the force density acting locally on the fluid particles. This force density will be used to model a van der Waals fluid described by the Navier-Stokes equations. 
The forcing term $\mathcal{F}_i$,  which appears in Eq.~\eqref{eq:be}, is expressed as a second order expansion in the discrete velocities $\mathbf{e}_i$\,:
\begin{equation}
\mathcal{F}_i=\omega_i\left[A+\mathbf{B}\cdot\mathbf{e}_i+\frac{1}{2}\mathbf{C}:(\mathbf{e}_i\mathbf{e}_i-\mathbf{I})\right]\,, 
\end{equation}
where A, $\mathbf{B}$, and $\mathbf{C}$ are functions of $\mathbf{F}$. 
With the choice 
\begin{subequations}
\begin{eqnarray}
 A & =& 0 \,, \\
 B_{\alpha} & = & \left(1-\frac{\Delta t}{2\tau}\right)F_{\alpha}\,, \\
 C_{\alpha\beta} & = & \left(1-\frac{\Delta t}{2\tau}\right)\left\{\mathrm{v}_\alpha^{*} F_{\beta}+ F_{\alpha}\mathrm{v}_\beta^{*}  + (1-T) \right. \nonumber \\  & \times & \left.[\mathrm{v}_{\alpha}^{*}\partial_\beta\rho+\mathrm{v}_{\beta}^{*}\partial_\alpha\rho+\partial_{\gamma}(\rho\mathrm{v}_\gamma^{*}\delta_{\alpha\beta})]\right\}\,, 
\end{eqnarray}
\end{subequations}
a second-order Chapman-Enskog expansion would lead to the mass and momentum conservation equations\,:
\begin{subequations}
\begin{eqnarray}
\partial_t\rho+\partial_{\alpha}(\rho\mathrm{v}_{\alpha}^{*}) & = &0\,, \label{cont}\\
 \partial_t(\rho\mathrm{v}_{\alpha}^{*})+\partial_{\beta}(\rho\mathrm{v}_{\alpha}^{*}\mathrm{v}_{\beta}^{*}) & = & -\partial_{\alpha}p^{i}+F_{\alpha} \nonumber \\
 & +
 &\partial_{\beta}\left[\eta(\partial_{\alpha}\mathrm{v}_\beta^{*}+\partial_{\beta}\mathrm{v}_\alpha^{*})\right] \,, \label{ns}  
\end{eqnarray}
\end{subequations}
where $p^{i}=\rho T$ is the ideal gas pressure, $\eta=\rho(\tau-\Delta t/2)$ is the shear viscosity, and
$F_{\alpha}=\partial_{\alpha}(p^{i}-p^{w})+\kappa \rho\partial_{\alpha}(\nabla^2\rho)$
are the Cartesian components of the force density $\mathbf{F}$. Here
$p^{w}=3\rho T/(3-\rho) -9\rho^2/8$ is the van der Waals pressure with the critical point at $\rho_c=1$ and $T_c=1$, and $\kappa$ is the parameter controlling the surface tension. The spatial derivatives in the expression of $F_{\alpha}$ are calculated using a second-order finite-difference scheme \cite{strikwerda}.
For the sake of clarity, we use the symbol $\mathbf{v}$ in place of $\mathbf{v}^\ast$ in the rest of the paper.
\textcolor{black}{The continuum assumption of hydrodynamics, as described by the continuity (\ref{cont}) and 
Navier-Stokes (\ref{ns}) equations, is valid as far as the Knudsen number $Kn$ is negligible. In our model it results to be $Kn=(\Delta s \tau) / (\Delta t L)$, $L$ being the system size. For our choice of parameters it is $Kn < 0.01$ thus making the continuum approximation reliable.}

We made use of a parallel approach implementing Message Passing Interface (MPI) to parallelize the code. The computational domain 
was divided into slices that were individually assigned to a particular task in the MPI communicator. Non-local operations (such as derivatives) were treated through the ghost-cell approach \cite{trobec2018introduction}. Simulations presented in the next Section were performed using $128$ CPU cores (on AMD Epyc Processors at the ReCaS Infrastructure), taking $72$ hours for each run.


\begin{figure}[t!]
 \centering
 \includegraphics[width=0.96\columnwidth]{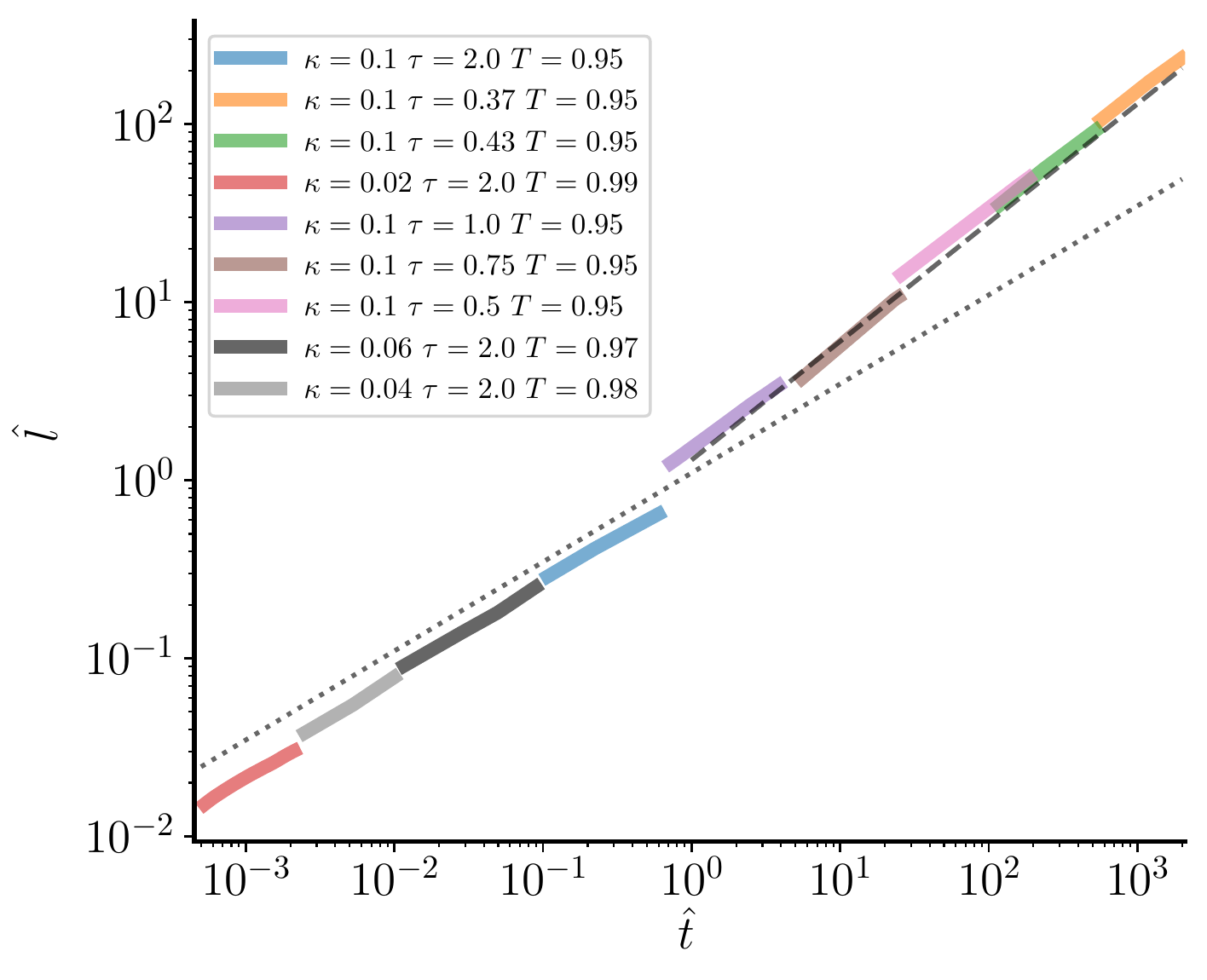}
 \caption{\textbf{Scaling.} Scaling of the dimensionless domain size $\hat{l}=L(t)/L_0$ as function of the dimensionless time $\hat{t}=(t-t_{int})/t_0$ (see text for details). The dotted and dashed black lines have slopes $1/2$ and $2/3$, respectively. In all the runs the interface width is $\xi\simeq4$.} \
 \label{fig:3_radius_scaling}
\end{figure}

\begin{figure*}[t!]
\centering
\includegraphics[width=1.94\columnwidth]{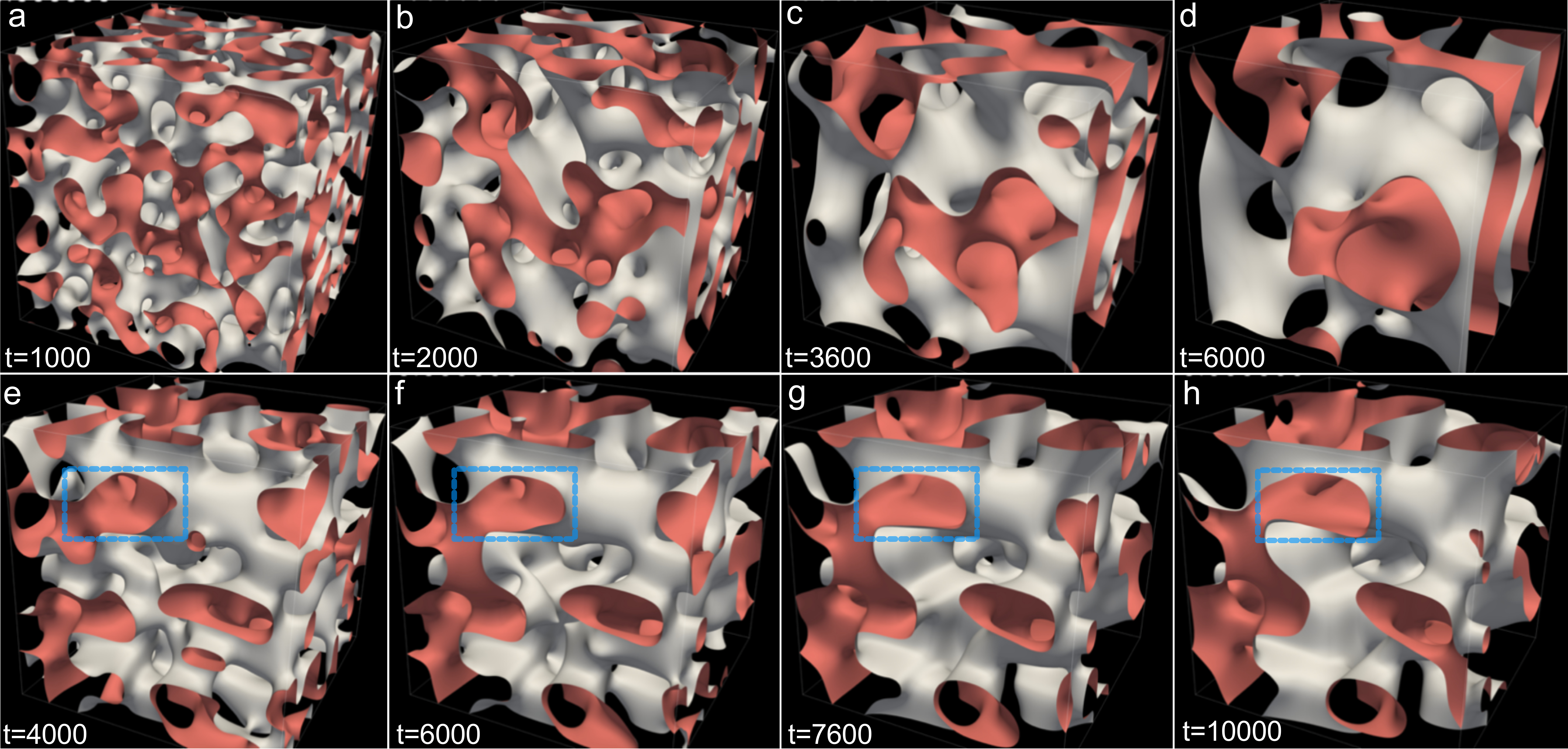}
\caption{\textbf{Morphological Characterization.} Snapshots of the system at consecutive times, showing the isosurfaces of $\rho_a$ for $\tau=0.5$ (upper row) and $\tau=2.0$ (lower row) with $T=0.95$ and $\kappa=0.1$. The cyan rectangles track the time evolution of a liquid-vapor interface during the \textcolor{black}{kinetic} growth regime (see Fig.~\ref{fig:1_radius}).}
\label{fig:2_snapshots}
\end{figure*}

\begin{figure}[ht!]
 \centering
 \includegraphics[width=0.95\columnwidth]{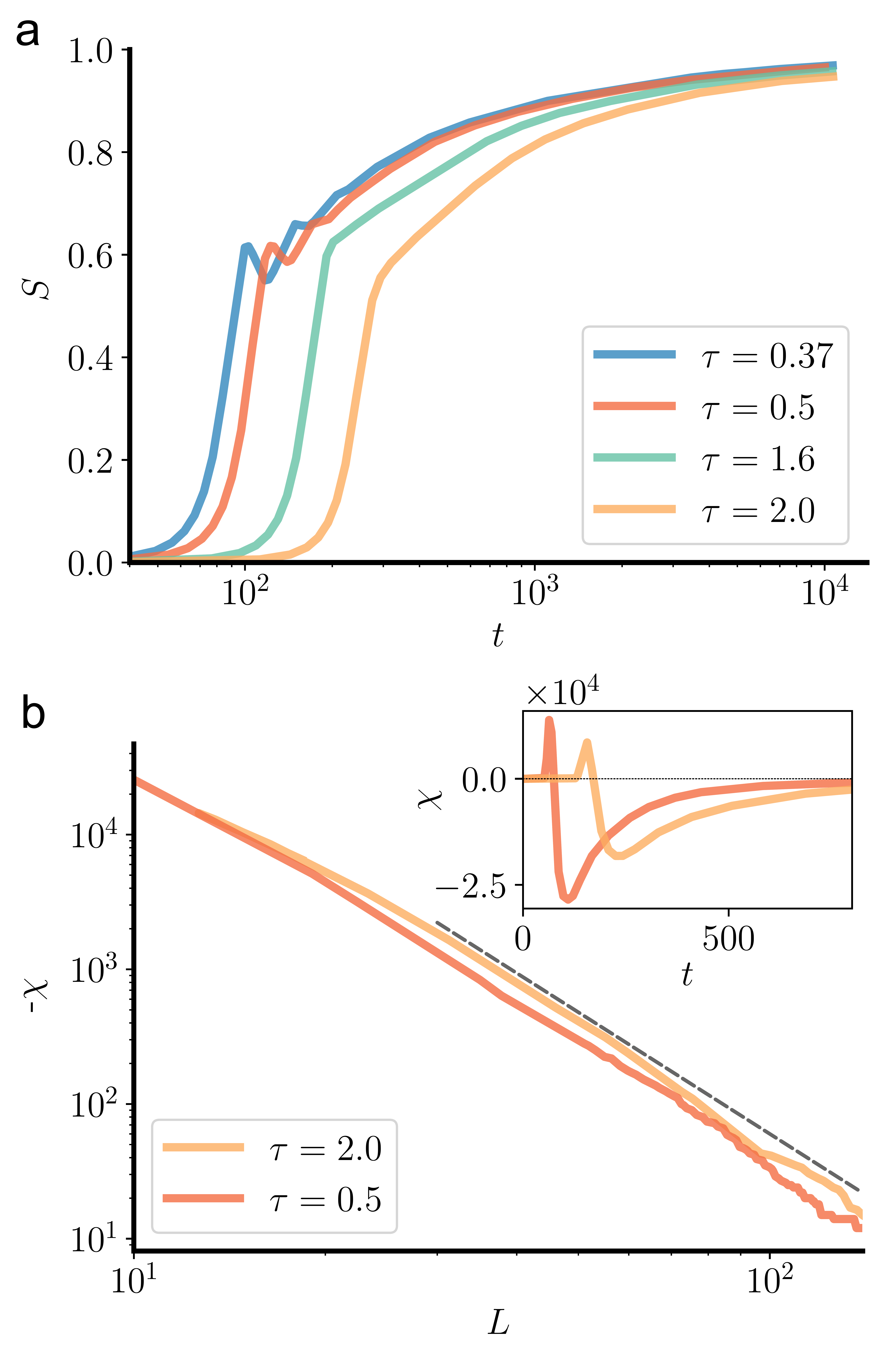}
 \caption{\textbf{Morphological Characterization.} \textbf{(a)} Separation depth $S(t)$ at $T=0.95$ and $\kappa=0.1$, for different values of $\tau$. \textbf{(b)} Euler characteristic $\chi$ as a function of mean domains size $L$ for $\tau=0.5$ (red curve) and $\tau=2.0$ (orange curve). The dashed curve has a slope of $-3$.  The inset of panel (b) shows the time evolution of $\chi$ at early times.}
 \label{fig:4_morphology}
\end{figure}
\section{Results}

The simulations were performed on three-dimensional lattices having $512$ nodes along each spatial direction with periodic boundary conditions. At $t = 0$, the systems were prepared in a symmetric disordered state
with small density fluctuations, namely $0.1\%$, around the mean value $\rho_a=(\rho_L+\rho_V)/2$, $\rho_L$ and $\rho_V$ being the coexisting densities of the liquid and vapor phases at the quenching temperature $T < T_c$, respectively.
The surface tension, which can be approximated as $\sigma=2 \kappa (\rho_L-\rho_V)^2/3 \xi$ \cite{angel1998},
where  $\xi=2 \sqrt{\,2 \kappa T /(1-T)\,}$ \cite{wagner07} is the interface width,
was changed by varying $\kappa$ and $T$. The viscosity was varied via the
relaxation time $\tau$. The results regarding the kinetics of the phase separation process,
the characterization of the morphology and the flow properties are presented below.


\subsection{Kinetics} 
In order to estimate the size of domains we measure the characteristic domain size $L(t)$ from the inverse of the first moment of the spherically averaged structure factor
 \begin{equation}\label{eq:Lt}
 L(t)=\pi \frac{\int S(k,t) dk}{\int k S(k,t) dk}\ ,
 \end{equation}
 where $k=|\mathbf{k}|$ is the modulus of the wave vector in the Fourier space. The structure factor is \cite{kendon1999,juo}
 \begin{equation}
 S(k,t)=\langle \hat{\rho}(\mathbf{k},t) \hat{\rho}(-\mathbf{k},t)\rangle_k\ ,
 \end{equation}
where $\hat{\rho}(\mathbf{k},t)$ is the spatial Fourier transform of the density difference
 $\rho-\rho_a$. 
The angle brackets denote an average over a
shell in $\mathbf{k}$-space at fixed $k$. 
 
 The domain growth is illustrated in Fig.~\ref{fig:1_radius} for
$T=0.95$ and $\kappa=0.1$  (these values ensure the interface width to be $\xi \simeq 4$), and
two values of $\tau$.  When $\tau = 0.5$ (the red curve), the observed
growth exponent is $\alpha_i = 2/3$, which is typical for the inertial hydrodynamic regime occurring at late times \cite{Furukawa1985,Furukawa2000}. When we increase the viscosity, the phase separation is delayed, as observed in the case with $\tau=2.0$ (the orange curve): after a long crossover,  the domain growth is characterized by the exponent $\alpha_d \simeq 1/2$ \cite{osborn1995,sofonea2004,xu2011,zhang2019}.

\begin{figure*}[htp]
\centering
\includegraphics[width=1.96\columnwidth]{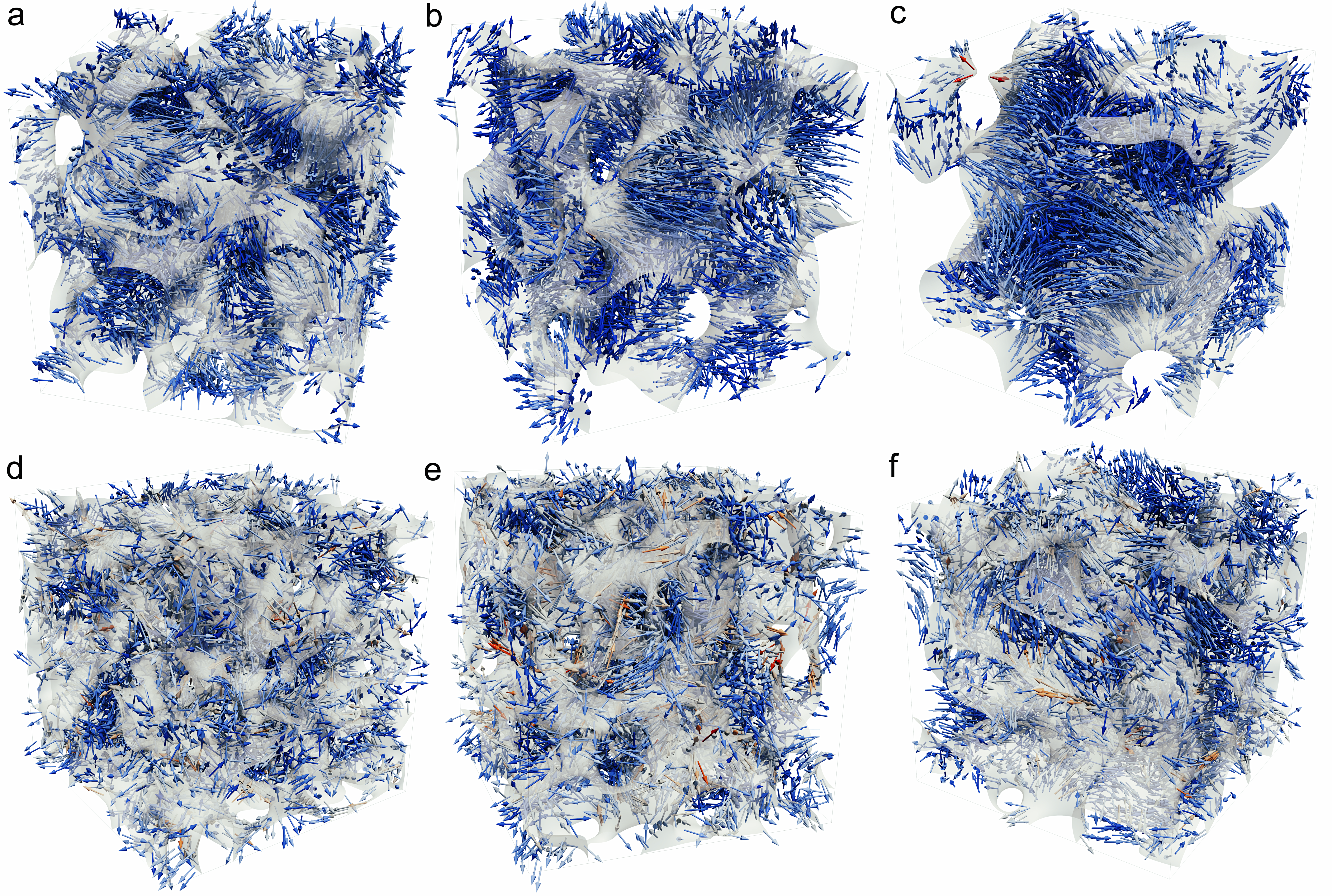}
\caption{\textbf{Flow properties.} Snapshots of the system at different times for $\tau=0.5$ (top row) and $\tau=2.0$ (bottom row){, with $T=0.95$ and $\kappa=0.1$}, showing the isosurfaces at $\rho=\rho_a$ with the velocity field next to the interface at times $t=6 \times 10^3, 9 \times 10^3, 1.4 \times 10^4$ (from left to right). }
\label{fig:vel_confs}
\end{figure*} 
Assuming that only capillary forces, viscous dissipation and fluid inertia are involved in liquid-vapor phase separation, as commonly assumed in previous works on fluid mixtures \cite{kendon1999,kendon2001}, the only control parameters are the viscosity $\eta$, the density $\rho$, and the surface tension $\sigma$. From these quantities, one can define a single length scale $L_0=\eta^2/\rho_a \sigma$ and a single time scale $t_0=\eta^3/\rho_a \sigma^2$ where we have used the average value $\rho_a$ as characteristic density.  After conducting several simulations with various values of $\kappa$, $\tau$ and $T$, keeping fixed the interface width, we fitted the resulted data sets $L(t)$ using the formula $a(t-t_{int})^{\alpha}$, where $a$, $t_{int}$ and $\alpha$ are free parameters. Plotting the simulations results using the dimensionless variables $\hat{t}=(t-t_{int})/t_0$ and $\hat{l}(t)=L(t)/L_0$ (see Fig.~\ref{fig:3_radius_scaling}), we can observe the growth exponents $\alpha_d = 1/2$ and $\alpha_i = 2/3$ over several time decades, with a crossover at $\hat{t} \simeq 1$. This result gives convincing evidence of the existence of the two aforementioned regimes with a transition between them, that seems to exclude the presence of an intermediate viscous regime for three-dimensional liquid-vapor systems, differently from what holds for binary fluid mixtures.


\subsection{Morphology} 
We now characterize the morphology of the two kinetic regimes observed. 
The differences between the patterns in the low and the high viscosity regimes may be observed by comparing the snapshots shown in Fig.~\ref{fig:2_snapshots}, where the two faces of the liquid-vapor interface are represented in white and coral color, respectively (see also movie 1 and movie 2  in the Supplementary Material). The snapshots were taken during the  time intervals $10^3 \leq t \leq6 \times 10^3$  (for $\tau=0.5$, first row), and $4 \times 10^3 \leq t \leq 10^4$ (for $\tau=2.0$, second row), when the growth exponents $\alpha_i=2/3$ and $\alpha_d=1/2$ are observed, respectively (see Fig.~\ref{fig:1_radius}). 
In the first case, the Laplace pressure difference between the inner and outer regions of a domain is expected to induce the formation of more spherical patterns, driving the fluid motion \cite{de2013capillarity}. The interfaces then evolve smoothly between isolated breakups and topological reconnections, corresponding to the breakage of fluid necks (see movie  1 in the Supplementary material). 
This mechanism, which is induced by curvature and controlled by inertia \cite{FURUKAWA1994}, seems to be the dominant one in this regime and is expected to endure until the phase separation process is completed. 
At higher viscosity, in the time regime where the exponent $\alpha_d$ is observed, which occurs at later times compared to the previous case, domains are already formed and are larger (see Fig.~\ref{fig:1_radius} and Fig.~\ref{fig:2_snapshots}). Topological reconnections are still visible, but much less frequent (see movie 2 in the Supplementary material). The slower growth appears in the delayed motion of interfaces (see the region tracked with cyan boxes in Fig.~\ref{fig:2_snapshots}).
To look into the average phase composition of the two-phase fluid,
we measure the distance of the single-phase density from the corresponding equilibrium value during the phase separation process. This is done by measuring the separation depth $S$ defined as
\begin{equation}
 S=\langle\frac{\rho(\mathbf{r})-\rho_a}{\rho_{eq}(\mathbf{r})-\rho_a}\rangle \ \ ,
\end{equation}
where $\rho_a$ is the initial mean density and brackets indicate volume averaging. Here $\rho_{eq}$ denotes the equilibrium density of the liquid phase, $\rho_L$, or the vapor phase, $\rho_V$, depending on the local density $\rho(\mathbf{r})$ as

\begin{equation}
\rho_{eq}(\mathbf{r})=
\begin{dcases}
\rho_{L},& \textrm{if } \rho(\mathbf{r}) > \rho_a \\
\rho_{V},& \textrm{if } \rho(\mathbf{r}) < \rho_a .
 \end{dcases}
\end{equation}
Figure \ref{fig:4_morphology}(a) shows the temporal evolution of the separation depth $S$ for various values of $\tau$, at $T=0.95$ and $\kappa=0.1$. For $\tau \leq 0.5$, the phase separation takes place in three recognizable steps. First, there is a time delay, when no detectable phase separation occurs. Then, at the onset of phase separation, the process is very rapid and the separation depth jumps to a value $\simeq 0.6$, meaning that the system reaches local equilibrium shortly after sharp interfaces are formed. Afterward, in the third stage, phase separation proceeds much more slowly, as the density gradients within the single-phase domains are very small, while the densities of the two phases across any interface change only very slowly in time, asymptotically approaching equilibrium at $S=1$. The behavior of $S$ changes when increasing $\tau$. At $\tau=2.0$, after the initial time delay, when domains start to form, the separation depth jumps to a value close to $0.5$ and then increases with values that stay always lower than in the cases with smaller values of $\tau$. This indicates that the system at higher viscosity relaxes more slowly to local equilibrium while phase separation takes place.

To better understand the different morphological behaviors in the low and high viscosity regimes, we use the Minkowski functionals \cite{serra1983}. These have been previously used also to characterize patterns in phase separation of complex fluids \cite{mecke1997,mecke1999,xu2011,MICHIELSEN2001461}. 
Since every continuous pattern can be decomposed in black and white convex subsets using a thresholding procedure \cite{MICHIELSEN2001461}, the first step is 
to assign a binary variable to each lattice site (black/white voxels)
by introducing a suitable density cut-off given by $\rho_a$ in our case.
For a cubic lattice, the four Minkowski functionals are the volume $V$, the surface area $A$, the mean breadth $B$ (directly proportional to the mean curvature \cite{MICHIELSEN2001461}) and the Euler characteristic $\chi$. 
The last quantity equals the number of regions of connected black voxels plus the number of completely enclosed regions of white voxels minus the number of tunnels, i.e. regions of white pixels piercing regions of connected black voxels.
The calculation of these morphological measures relies on the preliminary determination of the total number of black voxels $n_c$, black-white faces $n_f$, edges $n_e$ and vertices $n_v$ \cite{MICHIELSEN2001461}, so that one has
\begin{eqnarray}
V & = & n_c \, , \,\,\, A = -6n_c +2n_f \, , \,\,\,
B=(3n_c-2n_f+n_e)/2 \, , \nonumber \\
\chi & = & -n_c+n_f-n_e+n_v .
\end{eqnarray}

The inset of Fig.~\ref{fig:4_morphology}(b) shows the evolution of the Euler characteristics $\chi$ for two values of $\tau$.
 In both cases, after an early delay characterized by $\chi=0$ (a vanishing Euler characteristic indicates a highly connected structure with equal numbers of black and white domains), $\chi$ increases to values above zero, meaning that the number of domains with  $\rho>\rho_a$ increases. This marks the beginning of phase separation, corresponding to the first stage discussed at the beginning of the section. Immediately after, $\chi$ drops to negative values, meaning that the number of tunnels increases. After reaching a minimum value, $\chi$ starts to increase, signaling that the number of domains belonging to both phases starts to grow. For $\tau=2.0$ the process is delayed. In addition, the number of domains with $\rho>\rho_a$, as well as the number of tunnels, is lower at the beginning of the phase separation process. 
Following Refs.~\cite{mecke1997,mecke1999}, one can assume the following scaling behavior:
\begin{equation}\label{eq:scaling}
 V\sim 1 \ , \ \ A\sim L^{-1} \ , \ \ B\sim L^{-2}, \ \chi \sim L^{-3} \ .
\end{equation}
In Fig.~\ref{fig:4_morphology}(b) we show the Euler characteristic $\chi$ plotted against the mean domain size $L$ for $\tau=0.5$ and $\tau=2.0$. The curves show a good agreement with the proposed scaling. \textcolor{black}{Moreover, one can safely assume that, since the initial composition is symmetric, there is one continuous connected liquid phase (black voxels) and one continuous vapour phase (white voxels) throughout the phase separation process (see also the supplementary movies). Therefore, the value of $\chi$ will be determined by the number of vapour tunnels (i.e. regions of white voxels piercing regions of connected black voxels). The minimum value of $\chi$ in the inset of Fig.~\ref{fig:4_morphology}(b) corresponds to the maximum number of tunnels and to the onset of the phase separation. The pinch-off of each vapour tunnel (equivalent to the closure of liquid-phase domains \cite{Mcclure2020}) increases the value of $\chi$ by one unit. 
Consequently, the growth of $\chi$ is determined by the vapour tunnels pinch-offs and is connected to the growth of the characteristic length $L$ through the scaling in Eq.~\eqref{eq:scaling}. 
The difference between the low and the high viscosity cases in the inset of panel (b)
of  Fig.~\ref{fig:4_morphology} shows evidence of the lower rate of closures of pinch-offs when the viscosity
is higher.}



\subsection{Flow Field}
\begin{figure}[ht!]
\includegraphics[width=0.95\columnwidth]{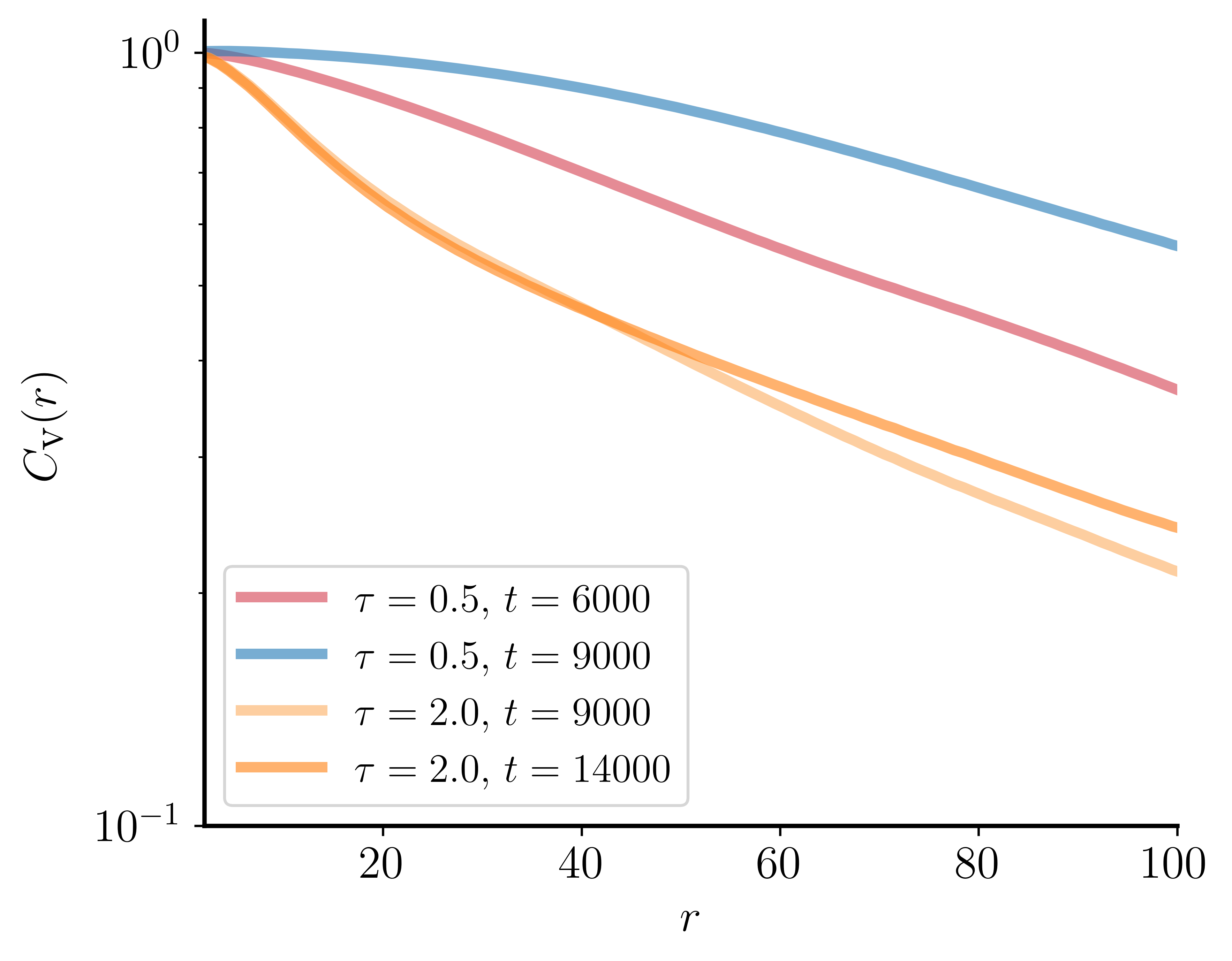}
\caption{\textbf{Flow properties.} Plot of the velocity correlation (Eq.~\eqref{eq:vel_corr}), for $\tau=0.5$ and $\tau=2.0$, at times corresponding to the configurations shown in Fig.~\ref{fig:vel_confs}(a,b,e,f). }
\label{fig:vel_corr}
\end{figure}
We now look more closely at the flow properties in the two regimes. Figure \ref{fig:vel_confs}
shows snapshots of the system with the isosurfaces $\rho=\rho_a$ and the superimposed velocity field colored by its magnitude. The snapshots are taken at the same times in the two regimes. Panels (a) and (e) have almost the same domain size $L_1 \simeq 80 $, as panels (b) and (f) for which the domain size is $L_2 \simeq 100$. A visual comparison shows that in the low viscosity (upper panels of Fig.~\ref{fig:vel_confs}) case coherent motion occurs inside domains with the same composition. Such correlated motion is lost when increasing the viscosity (see lower panels of Fig.~\ref{fig:vel_confs}). To characterize quantitatively the typical length scale of fluid velocity correlation, we measure the function 
\begin{equation}\label{eq:vel_corr}
C_{\textrm{v}}(r)=\frac{\langle\mathbf{v}(0,t)\cdot\mathbf{v}(r,t)\rangle}{\langle\mathbf{v}(0,t)^2\rangle} 
\end{equation}
where the brackets denote volume averaging.
The results for $\tau=0.5$ and $\tau=2.0$ at the times corresponding to the typical sizes $L_1$ and $L_2$ (in Fig.~\ref{fig:vel_confs}) are presented in Fig.~\ref{fig:vel_corr}. We observe an exponential decay in all the plotted cases, with $C_{\textrm{v}}\sim\exp(-r/\xi_\textrm{v})$. Two main differences appear between the two cases. At low viscosity the exponential decay is always slower than at high viscosity, meaning that the flow field correlates on a larger length scale. Moreover, in the inertial regime,  the correlation length increases from $\xi_v\simeq 93$ ($t=6 \times 10^3$) to $\xi_v\simeq133$ ($t=9 \times 10^3$)
while the typical size increases from $L_1$ to $L_2$, being always $\xi_v \gtrsim L$.
At high viscosity it results to be $\xi_v < L$ since the velocity correlation length scale grows from $\xi_v\simeq 64 (t=9 \times 10^3)$ to $\xi_v\simeq 86 (t=14 \times 10^3)$. The same loss of correlated motion has been observed and discussed in three-dimensional simulations of binary mixtures when going from low to high viscosity \cite{Pagonabarraga_2001}.

\section{Discussion and Conclusions}

In this work, using a LB model for liquid-vapor systems, we have studied the phase separation of a van der Waals fluid in three space dimensions \textcolor{black}{at fixed temperature}. 
\textcolor{black}{When the latent heat is negligible, the present isothermal model would be experimentally meaningful. This condition is realized at large length scales such that the thermal energy in the domains is much larger than the interfacial energy. In this case, a variation of the interface during phase separation will cause a negligible variation in the temperature. This condition applies to our model since the system size $L$ is much larger than the interface width $\xi$ being $L / \xi > 10^2$.}
We find evidence for \textcolor{black}{two} regimes with growth exponents $1/2$ and $2/3$ corresponding to high and low viscosity systems. By using rescaled length $\hat{l}$ and time $\hat{t}$ scales as in the case of binary fluids \cite{kendon2001}, we combined data from several simulations, with different values of viscosity, surface tension and temperature, to obtain a single curve of $\hat{l}$ as a function of $\hat{t}$.  We observe that the aforementioned growth regimes span several time decades with a crossover at scaled time $\hat{t} \simeq 1$. In our results, we do not find the existence of a viscous regime in the liquid-vapor phase separation, different from what is observed for binary mixtures \cite{kendon1999}. The morphological characterization of the observed growth regimes shows also some differences in the composition of liquid and vapor phases during the phase separation stages, and in the structure of the domains. 
The different scaling behaviors in the growth regimes are reflected in the morphological behavior.  This is witnessed by the 
scaling relation $\chi \sim L^{-3}$, connecting the Euler characteristic $\chi$ to the domain size $L$, which holds in both growth regimes.
A closer inspection of the flow field reveals a correlation on longer length scales in the inertial regime compared to the \textcolor{black}{kinetic} one. 

We hope our work is useful in clarifying the phase-separation process in liquid-vapor systems,  stimulating further investigations taking into account also the thermal effects. These latter are known to produce a rich phenomenology \cite{gonnellaPRE2007,zhang2019,xu2011},
 worth to be investigated in three spatial dimensions. 

\section*{Acknowledgements}
This work was possible thanks to the access to Bari ReCaS e-Infrastructure funded by MIUR through PON Research and Competitiveness 2007-2013 Call 254 Action I. We acknowledge funding from MIUR Project No. PRIN 2020/PFCXPE.
The work of AL was performed under the auspices of GNFM-INdAM
and supported by the National Centre for HPC, Big Data and Quantum Computing (Spoke 6, CN00000013).

\appendix*

\section{Movies Description}
\textcolor{black}{In this Section we provide a brief description of the movies accompanying the paper.
\begin{itemize}
\item{\textbf{Movie 1: Inertial Growth Dynamics}\\ The movie illustrates the time evolution of the liquid-vapor interface (represented by an isosurface of $\rho_a$), for $\tau=0.5$ and $T=0.95$. After a brief transient regime (see also Fig.1), during which small domains of the two phases are formed,   inertial growth with an exponent $\alpha_i=2/3$ commences (approximately at $t=700$). The interfaces evolve
smoothly between isolated breakups and topological reconnections, corresponding to the breakage of fluid necks. }
\item{\textbf{Movie 2: \textcolor{black}{Kinetic} Growth Dynamics}\\
The movie depicts the time evolution of the liquid-vapor interface for $\tau=2.0$ and $T=0.95$. Following the formation of small domains of the two phases, domain growth is once again propelled by breakups and topological reconnections. Conversely, during the \textcolor{black}{kinetic} regime (starting at $t \simeq 4000$), topological reconnections remain visible but occur much less frequently (see also the discussion of the time evolution of the Euler characteristic in the main text). The slower growth is evident in the delayed movement of interfaces.
}
\end{itemize}}
\bibliographystyle{unsrt}
\bibliography{biblio}
\end{document}